\documentclass[9pt,conference]{IEEEtran}
\IEEEoverridecommandlockouts

\usepackage{cite}
\usepackage{amsmath,amssymb,amsfonts}
\usepackage{algorithmic}
\usepackage{graphicx}
\usepackage{textcomp}
\usepackage{xcolor}
\usepackage{multirow}
\usepackage{authblk}
\usepackage{comment}
\usepackage{xcolor}
\usepackage{listings}
\usepackage{hyperref}
\usepackage{tikz}
\usepackage{color}
\usepackage[linesnumbered,ruled,vlined]{algorithm2e}
\usepackage{amsmath}

\hypersetup{
    colorlinks=true,
    linkcolor=blue,
    filecolor=magenta,      
    urlcolor=cyan,
    citecolor = magenta,
    pdfpagemode=FullScreen,
}

\definecolor{darkgreen}{rgb}{0.13, 0.55, 0.13} 

\def\BibTeX{{\rm B\kern-.05em{\sc i\kern-.025em b}\kern-.08em
    T\kern-.1667em\lower.7ex\hbox{E}\kern-.125emX}}
\begin{document}

\title{NSFlow: An End-to-End FPGA Framework with Scalable Dataflow Architecture for Neuro-Symbolic AI
}

\author{
Hanchen Yang$^{*1}$,
Zishen~Wan$^{*1}$,
Ritik Raj$^{1}$,
Joongun Park$^{1}$,
Ziwei Li$^{1}$,
Ananda Samajdar$^{2}$, 
\\
Arijit Raychowdhury$^{1}$,
Tushar Krishna$^{1}$
\\
\textit{$^1$Georgia Institute of Technology, Atlanta, GA ~~~ $^2$IBM Research, Yorktown Heights, NY} \\

\vspace{-20pt}
\thanks{

$^*$Equal Contributions. (hanchen@gatech.edu, zishenwan@gatech.edu)

This work was supported in part by CoCoSys, one of seven centers in JUMP 2.0, a Semiconductor Research Corporation program sponsored by DARPA.}
}

\vspace{-20pt}


\newcommand{\NAME}{NSFlow\space}
\newcommand{\GRAPH}{Operator Graph\space}
\newcommand{\DSE}{DAG\space}
\newcommand{\SA}{AdArray\space}
\newcommand{\todo}[1]{\textcolor{red}{TODO: #1}} 
\newcommand{\red}[1]{\textcolor{red}{#1}} 
\newcommand{\blue}[1]{\textcolor{blue}{#1}}
\newcommand*\circled[1]{\tikz[baseline=(char.base)]{
            \node[shape=circle,fill,inner sep=0.2pt] (char) {\textcolor{white}{#1}};}}
            
\newcommand{\ju}[1]{\textcolor{purple}{JU: #1}} 
\newcommand{\hy}[1]{\textcolor{red}{HY: #1}} 

\maketitle

\begin{abstract}

Neuro-Symbolic AI (NSAI) is an emerging paradigm that integrates neural networks with symbolic reasoning to enhance the transparency, reasoning capabilities, and data efficiency of AI systems. 
Recent NSAI systems have gained traction due to their exceptional performance in reasoning tasks and human-AI collaborative scenarios.
Despite these algorithmic advancements, executing NSAI tasks on existing hardware (e.g., CPUs, GPUs, TPUs) remains challenging, due to their heterogeneous computing kernels, high memory intensity, and unique memory access patterns.
Moreover, current NSAI algorithms exhibit significant variation in operation types and scales, making them incompatible with existing ML accelerators. These challenges highlight the need for a versatile and flexible acceleration framework tailored to NSAI workloads.


In this paper, we propose NSFlow, an FPGA-based acceleration framework designed to achieve high efficiency, scalability, and versatility across NSAI systems. \NAME features 
a design architecture generator that identifies workload data dependencies and creates optimized dataflow architectures, as well as a reconfigurable array with flexible compute units, re-organizable memory, and mixed-precision capabilities.
Evaluating across NSAI workloads, \NAME achieves 31$\times$ speedup over Jetson TX2, more than 2$\times$ over GPU, 8$\times$ speedup over TPU-like systolic array, and more than 3$\times$ over Xilinx DPU. \NAME also demonstrates enhanced scalability, with only 4$\times$ runtime increase when symbolic workloads scale by 150$\times$.
To the best of our knowledge, 
\NAME is the first framework to enable real-time generalizable NSAI algorithms acceleration, demonstrating a promising solution for next-generation cognitive systems.


\end{abstract}

\section{Introduction}
\label{sec:intro}



Neuro-Symbolic AI (NSAI) emerges as a promising paradigm toward achieving artificial general intelligence (AGI) and human-like fluid intelligence. Compared to deep neural networks (DNNs), NSAI exhibits superior performance in cognitive tasks such as human-like learning, reasoning, and logical thinking~\cite{mao2019neuro,han2019visual,mei2022falcon,yi2020clevrer,zhang2021abstract,shah2022knowledge,trinh2024solving,ibrahim2024special,wan2024towards_3}. NSAI synergistically combines neural approaches (e.g., DNNs) with symbolic representations (e.g., vectors, logics, graphs) to advance cognitive ability \cite{booch2021thinking,camposampiero2024towards,hersche2024probabilistic,hitzler2022neuro, hamilton2024neuro,wan2024towards_2,wan2024h3dfact}.



Despite its cognitive advantages, achieving real-time and efficient NSAI inference on resource-constrained devices presents significant challenges. These challenges stem from higher memory intensity, greater computational kernel heterogeneity, irregular memory access patterns, and underutilization of hardware resources. In our experiments, it takes $>$3~mins on NVIDIA desktop GPU to perform single reasoning task~\cite{hersche2023neuro}, underscoring the inefficiency of current solutions. 

Previous work has identified three main challenges of NSAI~\cite{wan2024towards}: 
\underline{\emph{First}}, \emph{high memory footprint}. 
NSAI systems heavily rely on vector-symbolic architectures (VSAs) that use vector operations to encode symbolic knowledge, resulting in large memory footprints (often tens to hundreds of MB) and making it impractical to be fully cached on-chip in hardware accelerators.
\underline{\emph{Second}}, \emph{heterogeneous compute kernels}. Beyond neural networks, NSAI workloads incorporate diverse computations of varying sizes, such as vector convolutions, element-wise operations, and logical reasoning. These exhibit low data reuse, low compute array utilization, and limited parallelism, leading to inefficiencies on GPUs and TPUs.
\underline{\emph{Third}}, \emph{critical path dependency}. Symbolic reasoning often depends on outputs from neuro-perceptual modules, extending the critical path during cognitive inference and causing underutilization of traditional accelerators.

With the growing demand for scalable dataflow and architecture solutions for NSAI, FPGAs present an ideal platform due to their customizability, flexible memory management, and reconfigurability to adapt to evolving NSAI workloads. Previous work has demonstrated the potential of FPGAs for accelerating ML workloads~\cite{wang2021autosa, wei2017automated, li2020survey, chen2024understanding, chen2024comprehensive, zeng2024flightllm}. However, FPGA deployment remains challenging for NSAI algorithms due to the complexity of organizing on-chip resources and limited memory capacity \cite{huang2021fpga, yin2018high, xu2024llamaf}.

To address these challenges, we identify unique opportunities to enhance NSAI acceleration efficiency and propose NSFlow, a scalable FPGA-based dataflow architecture design automation framework. To the best of our knowledge, NSFlow is the \emph{first} automated end-to-end solution for accelerating and deploying generic NSAI workloads.
\NAME features a frontend subsystem with \emph{dataflow architecture generator} that includes NSAI execution trace extraction, dataflow graph generation, and a two-phase design space co-exploration strategy, and a backend subsystem with \emph{flexible neuro-symbolic hardware architecture} with adaptive array folding, reconfigurable memory partitioning, and efficient heterogeneous storage.
By integrating its frontend and backend, NSFlow delivers efficient and scalable acceleration for NSAI workloads. Specifically, it identifies data dependencies, explores design space options, and generates optimized dataflow architectures tailored for FPGA deployment.



This paper, therefore, makes the following contributions:
\begin{enumerate}
    \item An end-to-end FPGA design automation framework for accelerating and deploying generic NSAI workloads. 
    \item A design generator that (i) identifies workload-specific data dependencies using a self-generated dependency graph tailored for vector-symbolic-based NSAI algorithms and (ii) derives an optimal dataflow architecture through a novel design space co-exploration strategy.
    \item A hardware architecture featuring a flexible neuro-symbolic systolic array, an efficient SIMD unit, reorganizable on-chip memory, and support for mixed-precision computations.
\end{enumerate}

\section{Neuro-Symbolic AI and Characterization}
\label{sec:background}



This section presents NSAI algorithms with key kernels (Sec.~\ref{subsec:nsai}), and analyzes their workload characteristics (Sec.~\ref{subsec:profile}).

\begin{table*}[t!]
\centering
\caption{\textbf{Neurosymbolic models.} 
Selected neurosymbolic AI workloads for analysis, representing a diverse of application scenarios.}
\Huge
\renewcommand*{\arraystretch}{1.15}
\resizebox{\linewidth}{!}{%
\centering
\setlength\tabcolsep{5pt}
\begin{tabular}{cc|c|c|c|c}
\hline
\multicolumn{2}{c|}{\textbf{\begin{tabular}[c]{@{}c@{}}Representative Neuro- \\ Symbolic AI Workloads\end{tabular}}}      
& \textbf{\begin{tabular}[c]{@{}c@{}}Neuro-Vector-Symbolic \\ Architecture (NVSA)~\cite{hersche2023neuro}\end{tabular}}                                                              & \textbf{\begin{tabular}[c]{@{}c@{}}Multiple-Input-Multiple-Output \\ Neural Networks (MIMONet)~\cite{menet2024mimonets}\end{tabular}}                                                         & \textbf{\begin{tabular}[c]{@{}c@{}}Probabilistic Abduction via Learning Rules\\ in Vector-symbolic Architecture (LVRF)~\cite{hersche2024probabilistic}\end{tabular}}                                                                                           & \textbf{\begin{tabular}[c]{@{}c@{}}Probabilistic Abduction and\\ Execution Learner (PrAE)~\cite{zhang2021abstract}\end{tabular}}                                                                                                                                                          \\ \hline
\multicolumn{1}{c|}{\multirow{2}{*}{\textbf{\begin{tabular}[c]{@{}c@{}}Compute \\ Pattern\end{tabular}}}} & \textbf{Neuro}                                                                                                                                                                                                                                                                                                                 & CNN                                                                                                                                                                                                                     & CNN/Transformer                                                                                                                                                                      & CNN                                                                                                                                                                                       & CNN                                                                                                                                                                                                                                               \\ \cline{2-6} 
\multicolumn{1}{c|}{}                                                                                         & \textbf{Symbolic}                                                                                                                                                                                                                                                                                                           &     VSA binding/unbinding (Circular Conv)
&           VSA binding (Circular Conv) 
&  VSA binding/unbinding (Circular Conv)
& Probabilistic abduction                                                                                 \\ \hline
\multicolumn{1}{c|}{\multirow{3}{*}{\textbf{\begin{tabular}[c]{@{}c@{}}Application\\ Scenario\end{tabular}}}}  & \textbf{Use Case}                                                                                       & \begin{tabular}[c]{@{}c@{}}Spatial-temporal and abstract reasoning\end{tabular}                                                                                                                             & \begin{tabular}[c]{@{}c@{}}Multi-input simultaneously processing\end{tabular}   & \begin{tabular}[c]{@{}c@{}}Probabilistic reasoning, OOD data processing\end{tabular}                                                           & \begin{tabular}[c]{@{}c@{}}Spatial-temporal and abstract reasoning\end{tabular}                                                                                                                                                           \\ \cline{2-6} 
\multicolumn{1}{c|}{}                                                                                         & \textbf{\begin{tabular}[c]{@{}c@{}}Advantage \\vs. Neural\end{tabular}}                                                                                                                 & \begin{tabular}[c]{@{}c@{}} Higher joint representation efficiency,\\ Better reasoning capability, Transparency\end{tabular}                                                                   & \begin{tabular}[c]{@{}c@{}}Higher throughput, Lower latency, \\ Compositional compute, Transparency\end{tabular}                        & \begin{tabular}[c]{@{}c@{}}Stronger OOD handling capability, One-pass\\ learning, Higher flexibility, Transparency\end{tabular} & \begin{tabular}[c]{@{}c@{}}Higher generalization, Transparency, \\Interpretability, Robustness\end{tabular} \\ \cline{2-6} 
\hline
\end{tabular}
}

\begin{minipage}[b]{\linewidth}
    \centering
    \includegraphics[width=\columnwidth]{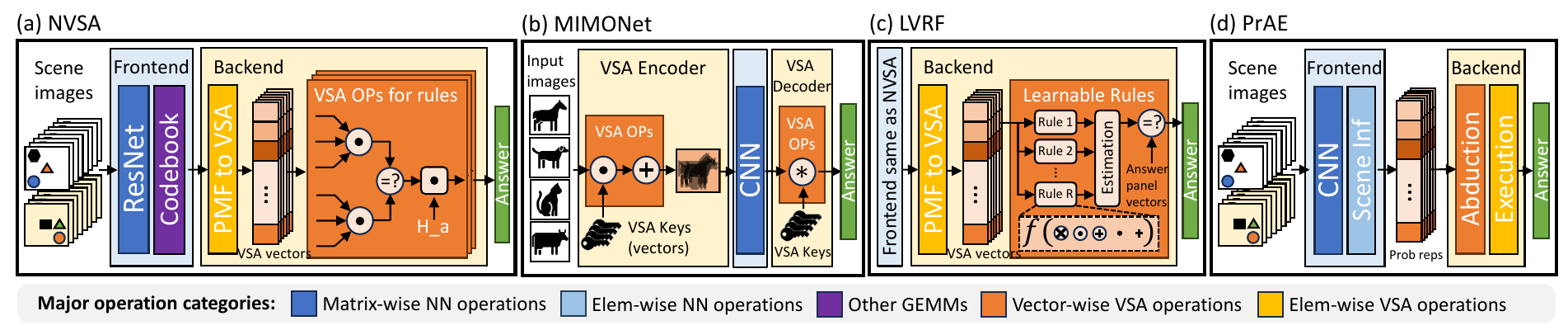}
\end{minipage}%

\label{tab:selected_model}
\vspace{-10pt}
\end{table*}


\subsection{Neuro-Symbolic AI Algorithm}
\label{subsec:nsai}
Neurosymbolic AI synergistically integrates \emph{learning capability} of neural networks with \emph{reasoning capability} of symbolic AI, offering data-efficient learning and transparent, logical decision-making beyond traditional DNNs. 
\circled{1} \textbf{Neural system.} The process begins with a neural module that handles perception tasks by interpreting sensory data and generating meaningful scene and object representations, providing essential inputs for reasoning. 
\circled{2} \textbf{Symbolic system.} These features are then passed to the symbolic system for reasoning, enhancing explainability and reducing reliance on extensive training data by leveraging established models of the physical world (e.g., rules and coded knowledge). This step integrates learned neural network knowledge with symbolic rules, allowing the system to both learn from new data and reason logically based on existing knowledge. The outputs of symbolic reasoning are used for decision-making, and response or action generation.

Tab.~\ref{tab:selected_model} highlights four representative neuro-symbolic workloads: NVSA\cite{hersche2023neuro} for spatial-temporal reasoning, MIMONet for multi-input processing~\cite{menet2024mimonets}, LVRF for probabilistic abduction~\cite{hersche2024probabilistic}, and PrAE for abstract reasoning tasks~\cite{zhang2021abstract}. These workloads demonstrate superior reasoning capabilities and represent a promising paradigm for human-like intelligence. These approaches integrate CNNs for neuro and vector-symbolic architectures (VSAs) for symbolic processing. 

A key VSA operation is the blockwise \textit{circular convolution} that combines two vectors in a way that preserves the information from both, making it suitable for representing composite symbols. Mathematically, the circular convolution of two vectors \( \mathbf{A} \) and \( \mathbf{B} \) (each of dimension \( N \)) generates vector \( \mathbf{C} \) as $C[n] = \sum_{k=0}^{N-1} A[k] \cdot B[(n-k) \mod N]$ where each element of \( \mathbf{C} \) is obtained by multiplying the elements of \( \mathbf{A} \) with the circularly shifted elements of \( \mathbf{B} \), and then summing up. Circular convolution has commutativity and associativity properties, making it particularly effective in hierarchical reasoning tasks where manipulating structured information is critical.

\begin{figure}[t!]
\centering
\vspace{-0.03in}
\includegraphics[width=\columnwidth]{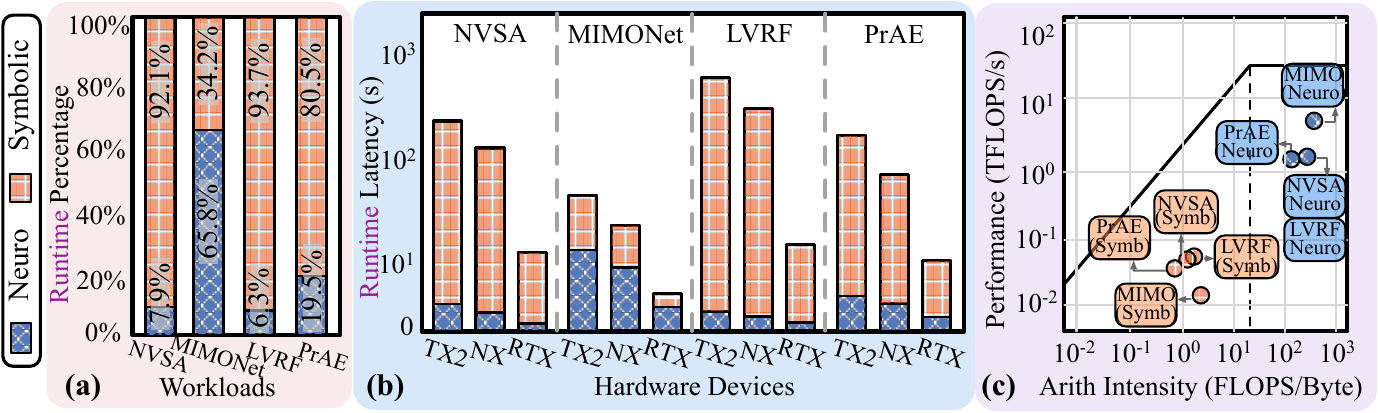}
\vspace{-0.2in}
        \caption{\textbf{End-to-end neuro-symbolic runtime and roofline characterization.} \textbf{(a)} \textnormal{Benchmark neuro-symbolic models on CPU+GPU system, showing symbolic may serve as system bottleneck. \textbf{(b)} Benchmark on Coral TPU, TX2, NX, and 2080Ti GPU, showing that real-time performance cannot be satisfied. \textbf{(c)} Roofline of RTX 2080Ti GPU, indicating symbolic memory-bounded.}} 
        \label{fig:profiling_latency_e2e}
        \vspace{-12pt}
\end{figure}

\subsection{Neuro-Symbolic AI Workload Characterization}
\label{subsec:profile}
To understand the real-device efficiency of neuro-symbolic AI workload, recent work~\cite{cogsys} profiles four representative models as elaborated in Tab.~\ref{tab:selected_model} on Coral edge TPU (4~W), Jetson TX2 (15~W), Xavier NX (20~W), and RTX 2080Ti (250~W), respectively.

\textbf{End-to-end latency breakdown.}
Fig.~\ref{fig:profiling_latency_e2e}\textcolor{blue}{a} and Fig.~\ref{fig:profiling_latency_e2e}\textcolor{blue}{b} present the end-to-end latency breakdown of neuro-symbolic workloads, highlighting three key observations: 
(1) \textit{The real-time performance cannot be satisfied} across devices. Even with additional compute resources to reduce NN runtime, the substantial overhead from symbolic reasoning prevents real-time execution.
(2) \textit{Symbolic operations dominate runtime.} For instance, symbolic modules account for 87\% of NVSA total runtime while contributing only 19\% of its total FLOPS, suggesting that symbolic operations are not efficiently handled by GPUs/TPUs.
(3) \textit{Symbolic reasoning computation lies on the critical path} as its computation depends on outputs from the neural modules. 


\textbf{System Roofline Analysis.} Fig.~\ref{fig:profiling_latency_e2e}\textcolor{blue}{c} employs the roofline model of RTX 2080Ti GPU version to quantify the neurosymbolic workloads. We observe that \textit{symbolic modules are memory-bounded while neuro modules are compute-bounded}. This is mainly due to symbolic operations requiring streaming vector elements, increasing the memory bandwidth pressure and resulting in hardware underutilization.



\section{NSFlow Overview}
\label{sec:overview}




\begin{figure}[t!]
\vspace{-10pt}
\centering
\includegraphics[width=.9\columnwidth]{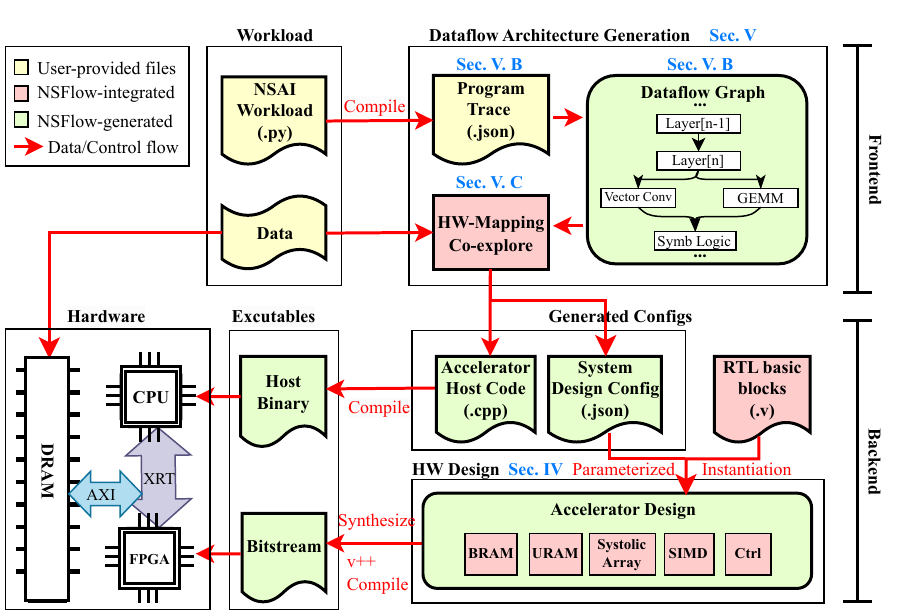}
        \vspace{-0.05in}
        \caption{ \textbf{NSFlow Overview.}}
        \label{fig:overview}
        \vspace{-15pt}
\end{figure}

\label{sec:accelerator}

NSFlow is an end-to-end framework 
that identifies data dependency, explores the design space, and generates an optimal dataflow architecture design for FPGA deployment tailored to a given NSAI workload. Fig.~\ref{fig:overview} (a) shows an overview of the proposed framework, divided into frontend and backend.

\subsection{\NAME Frontend}
\label{subsec:frontend}
The \emph{Design Architecture Generator (DAG)} is the core component of the frontend operating on the host side. The DAG module begins by extracting an execution trace from the user-provided workload. It then generates a dataflow graph specifically designed for VSA-based NSAI workloads, capturing operator-level specifications, runtime, memory functions, and their data dependencies. This dataflow graph is used for co-exploration of dataflow architecture through a novel two-phase algorithm. The first phase identifies the optimal system design configuration for the FPGA accelerator, while the second determines an efficient (or near-optimal) reconfiguration and mapping scheme. These configurations are specified in the accelerator host code, enabling the CPU to invoke device kernels via the XRT API. After compilation, the CPU executes the host binary to schedule operations on the FPGA.



\subsection{\NAME Backend}
\label{subsec:backend}
The \NAME backend includes a pre-define accelerator template comprising several essential components: BRAM blocks for flexible on-chip memory, adaptive Systolic Array for parallel neuro and symbolic operations, SIMD unit for element-wise, vector reduction and scalar operations, and control logic for task scheduling on hardware level. 
These components are parameterized using the system design configuration file generated by the frontend, enabling the instantiation of an optimized microarchitecture based on workload characterization. NSFlow then synthesizes and compiles the RTL into an executable bitstream for deployment.
During real-time inference, the CPU executes the host binary code to run FPGA kernels and manages off-chip memory transactions through AXI interfaces.


We will present the \NAME design in a bottom-up manner, starting with the backend flexible neuro-symbolic hardware architecture (Sec.~\ref{sec:arch}) and followed by the frontend graph and dataflow architecture generators (Sec.~\ref{sec:dse}).

\section{NSFlow Backend: Flexible Hardware Architecture}
\label{sec:arch}





\begin{figure*}[t!]
\centering
\includegraphics[width=.75\textwidth]{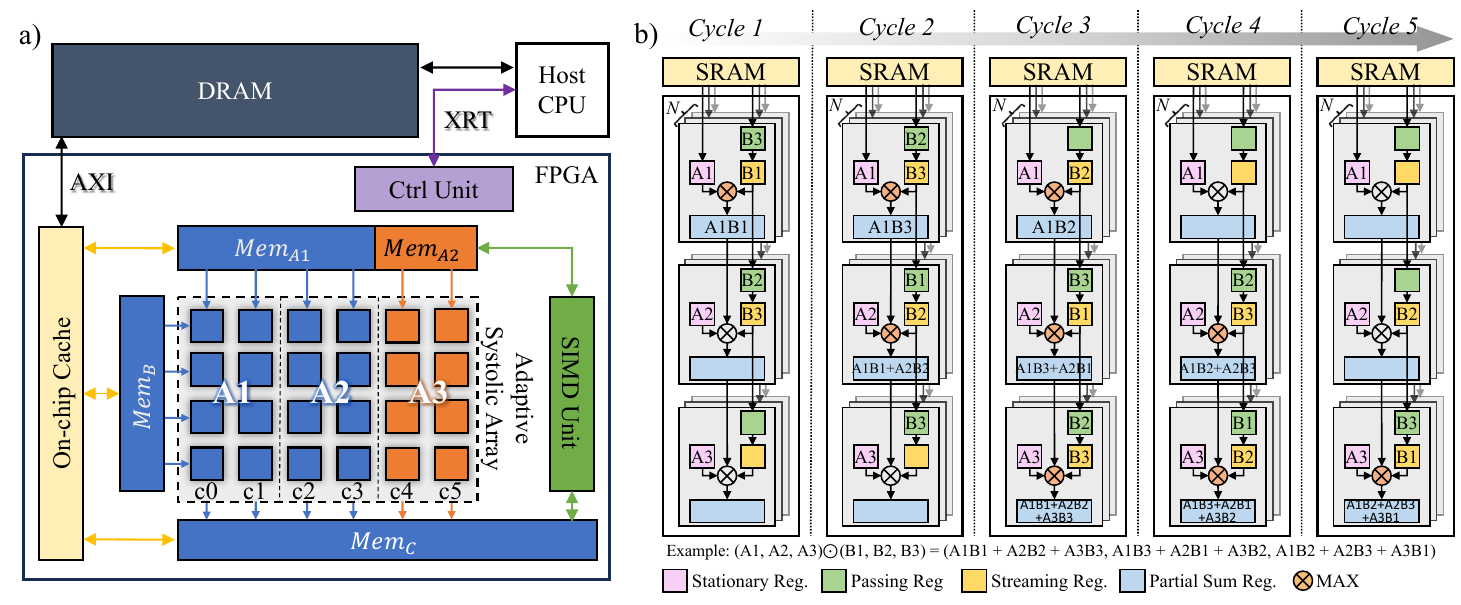}
        \vspace{-5pt}
        \caption{\textbf{NSFlow Hardware Architecture.}
        }
        \label{fig:arch}
        \vspace{-10pt}
\end{figure*}

This section first presents an overview of the \NAME hardware architecture (Sec.~\ref{subsec:hw_overview}), then walks through our design featuring an \textit{NS-adaptive systolic array} (Sec.~\ref{subsec:array}), \textit{Re-organizable on-chip memory} (Sec.~\ref{subsec:mem}), \textit{Adaptive mixed precision computation} (Sec.~\ref{subsec:precision}), an \textit{Efficient custom SIMD unit} (Sec.~\ref{subsec:simd}).


\subsection{Overview of NSFlow Hardware Architecture} 
\label{subsec:hw_overview}
Fig.\ref{fig:arch}(a) exhibits the hardware architecture of NSFlow. It consists of a uniquely designed NSAI-workload-adaptive Systolic Array for NN and Vector-symbolic operations, a SIMD unit for reductions, element-wise operations, flexibly arranged on-chip RAM blocks, and a control unit for kernel scheduling and memory transactions. NSFlow has pre-defined RTL of all the above blocks with scaling parameters subject to the design configuration generated from \DSE for optimal execution.


\subsection{Adaptive Systolic Array (AdArray)} 
\label{subsec:array}

Inspired by~\cite{cogsys}, we implement an adaptive systolic array design (AdArray) to maximize efficiency for NSAI inference.

\textbf{Adaptive array folding.} \SA can run both NN ops and vector-symbolic circular convolution, the two most dominating components in our targeted NSAI workload, on its arbitrary portions (sub-arrays) simultaneously to maximize parallelism and utilization. Each sub-array is either combined with its adjacent one to operate NN ops, or singularly running vector operations like circular convolution in the symbolic binding process. In Fig.~\ref{fig:arch}(a) we showcase the design with a mini 4$\times$6 systolic array, which is split into 3 sub-arrays with each ranging 2 columns (c0 and c1 for A1, c2 and c3 for A2, c4 and c5 for A3). In this case, A1 and A2 are combined together to perform NN ops, while in A3 each column is running vector-symbolic operations. 


\textbf{Efficient vector-symbolic circular convolution streaming.} Traditional TPU's systolic array is extremely inefficient for circular convolution operations with heavy memory transactions and low parallelism due to non-ideal spatial and temporal mapping. Fig.~\ref{fig:arch}(b) showcases how a single column in our design performs vector-symbolic operation with a 3-element circular convolution example. The first vector $\boldsymbol{A}$ is held in stationary registers, while the second vector $\boldsymbol{B}$ is streamed from SRAM. The MAC unit processes the data from both stationary and streaming registers, adding it to the partial product received from the PE above. A passing register temporarily stores the streaming input for a cycle before it moves to the streaming register. This value is transferred to the passing register of the next PE in the following cycle. The procedure is repeated until the final circular convolution outputs. Unlike traditional Systolic Arrays, each PE uses an extra register named Passing Reg at the top of one of the input ports to cause a 1-cycle streaming pace mismatch between the two input vectors $\boldsymbol{A}$ and $\boldsymbol{B}$, thus enabling circular convolution operations. Note that to enable this type of streaming, each PE needs to have one extra vertical input port connected with its above PE's previous right output port as depicted in Fig.~\ref{fig:arch}(b). When performing NN operations, the Passing Register is bypassed via multiplexer, and the horizontal connections in the sub-array are again established to enable weight and input passing as in a traditional Systolic Array. 

\textbf{Two-level flexibility.} Our efficient systolic array design also benefits from extraordinary flexibility at both design level and kernel level. At design level, our \DSE decides the array size and its memory size (Sec.~\ref{subsec:mem}), as well as the number of sub-arrays that best fits the overall workload characteristic (Sec.~\ref{sec:dse}); At kernel level, the array are reconfigured to an optimal folding scheme at runtime, maximizing utilization and parallelism for the NN and vector-symbolic operations.


\subsection{Re-Organizable On-Chip Memory}
\label{subsec:mem}
The profiled NSAI workloads feature heavy memory usage and versatile computing kernels, thus we design a flexible memory system to enable smooth executions and transactions with limited FPGA on-chip memory resource ($\sim$36MB on ZCU104), which features \circled{1} \textbf{Re-organizable memory partition}, \circled{2} \textbf{Adaptive memory size}, and \circled{3} \textbf{Efficient heterogeneous storage} to fully exploit FPGA's potential and maximize memory efficiency for NSAI workloads. 

As shown in Fig.~\ref{fig:arch}, the on-chip memory system consists of three memory blocks $Mem_A$, $Mem_B$, and $Mem_C$, an on-chip cache, and a memory bus for off-chip transactions. $Mem_A$, $Mem_B$, and $Mem_C$ are all double-buffered memories to enable seamless read and write among off-chip memory and the systolic array. 
\circled{1} $Mem_A$ is partitioned into two chunks - $Mem_{A1}$ and $Mem_{A2}$ - to simultaneously load NN layers and vector data for the corresponding sub-array in \SA. When performing NN operations or vector operations singularly, the two memory chunks can be merged into one at runtime for better performance and simpler control. $Mem B$ works as the IFMAP buffer in normal systolic arrays which feeds data to the horizontal inputs of \SA only for NN processing. $Mem_C$ stores the outputs from \SA and the SIMD unit which are either read by the compute units, or written to $Mem_A$/${Mem_B}$ or off-chip DRAM. The on-chip cache buffers intermediate results for the 3 memory blocks. 
\circled{2} The sizes of all above memory components will be defined by \DSE based on workload's characteristics and dataflow. 
\circled{3} In real FPGA deployment, $Mem_A$, $Mem_B$, and $Mem_C$ are comprised of numerous 18KB BRAM blocks for maximum configurability, and on-chip cache is built with URAM considering its large capacity (288KB per block). Small registers and buffers in compute element use LUTRAMs for fast and dynamic access.


\subsection{Adaptive Compute for Mixed Precision} 
\label{subsec:precision}
To improve computing efficiency and save on-chip memory usage, \NAME supports mixed precisions ranging from FP16/8 to INT8/4 in different components of the workload specified by user at frontend. \DSE employs compute units adaptive to various precisions. 
In NVSA for example, NN and Symbolic operations are quantized to INT8 and INT4 respectively, thus the multipliers in \SA and the SIMD support both precisions with sufficient leverage of DSP units \cite{langhammer2020high}. Low-precision additions are handled by LUT for fast outcome.


\subsection{Efficient Custom SIMD Unit} 
\label{subsec:simd}
\NAME incorporates a custom SIMD unit to efficiently perform vector reductions, element-wise operations, etc., with fluid data transfer between the output of the \NAME array and the input SRAM for successive executions. It comprises multiple processing elements (PEs), each equipped with compact logic circuits (i.e., sum, mult/div, exp/log/tanh, norm,
softmax, etc.) to handle vector operations or optimized sparse computations on mixed level of quantized data.

\section{\NAME Frontend: Dataflow Architecture Generation}
\label{sec:dse}





In the frontend, we implement a \textit{Dataflow Architecture Generator (DAG)} that first builds a dataflow graph based on operation trace extracted from the workload, then generates an optimal (or sub-optimal) dataflow architecture design, defined by a design configuration file for instantiating hardware modules, and a host code for the CPU to schedule accelerator kernels.

This section first identifies the NSAI dataflow challenges (Sec.~\ref{subsec:challenges}). Then we 
illustrates the process to generate \textit{operation graph} and subsequently \textit{dataflow graph} (Sec.~\ref{subsec:data_dependency}). Finally, we discuss how \textit{\DSE}searches for an optimal architectural design and mapping based on the \textit{dataflow graph} (Sec.~\ref{subsec:dse}). 


\subsection{NSAI Dataflow Challenges}
\label{subsec:challenges}

We identify three main NSAI dataflow challenges (Fig.~\ref{fig:dataflow}(b)). \underline{First}, the sequential execution and frequent interactions of neural and symbolic components results in increased latency and low system throughput. \underline{Second}, the heterogeneous neural and symbolic kernels lead to low compute array utilization and efficiency of ML accelerators. \underline{Third}, heavy memory transactions exhibited in both components can cause large communication latency, which is even more challenging in FPGA deployment. 

\DSE perfectly addresses the above challenges by \underline{first}, identifying data dependencies through dataflow graph to fully exploit parallelism opportunities among NN and symbolic operations with featured HW architecture; \underline{second}, balancing NN and symbolic operations on our \SA, empowered by both \textit{design-level flexibility} and \textit{kernel-level flexibility} to achieve maximum utilization; \underline{third}, configuring memory units adaptively to best-fit workload's memory usage, thus eliminating unnecessary transactions and stalls.


\subsection{Data Dependency Identification}
\label{subsec:data_dependency}


\begin{lstlisting}[caption={Neuro-Vector-Symbolic
Architecture Profiling Result}, label={lst:resnet_model}]
graph():
    ...
    // Neuro Operation - CNN (Resnet18)
    %relu_1[16,64,160,160] : call_module[relu](args = (%bn1[16,64,160,160]))
    %maxpool_1[16,64,160,160] : call_module[maxpool](args = (%relu_1[16,64,160,160]))
    %conv2d_1[16,64,160,160] : call_module[conv2d](args = (%maxpool_1[16,64,160,160]))
    ...
    // Symbolic Operations
    // Inverse binding of two block codes vectors by blockwise cicular correlation
    %inv_binding_circular_1[1,4,256] : call_function[nvsa.inv_binding_circular](args = (%vec_0[1,4,256], %vec_1[1,4,256]))
    %inv_binding_circular_2[1,4,256] : call_function[nvsa.inv_binding_circular](args = (%vec_3[1,4,256], %vec_4[1,4,256]))
    // Compute similarity between two block codes vectors
    %match_prob_1[1] : call_function[nvsa.match_prob](args = (%inv_binding_circular_1[1,4,256], %vec_2[1,4,256]))
    // Compute similarity between a dictionary and a batch of query vectors
    %match_prob_multi_batched_1[1]: call_function[nvsa.match_prob_multi_batched](args = (%inv_binding_circular_2[1,4,256], %vec_5[7,4,256]))
    %sum_1[1] : call_function[torch.sum](args = (%match_prob_multi_batched_1[1]))
    %clamp_1[1] : call_function[torch.clamp](args = (%sum_1[1]))
    %mul_1[1] : call_function[operator.mul](args = (%match_prob_1[1], %clamp_1[1]))
    ...
\end{lstlisting}




\textbf{Program trace.} \NAME first extracts an execution trace from input program through compilation, and pre-process the file to be ready for later dataflow graph generating. Listing \ref{lst:resnet_model} exhibits a snapshot taken from NVSA program trace, with representative kernels for Neural and Symbolic parts showing data dependencies.

\begin{figure}[t]
\centering
\vspace{-10pt}
\includegraphics[width=\columnwidth]{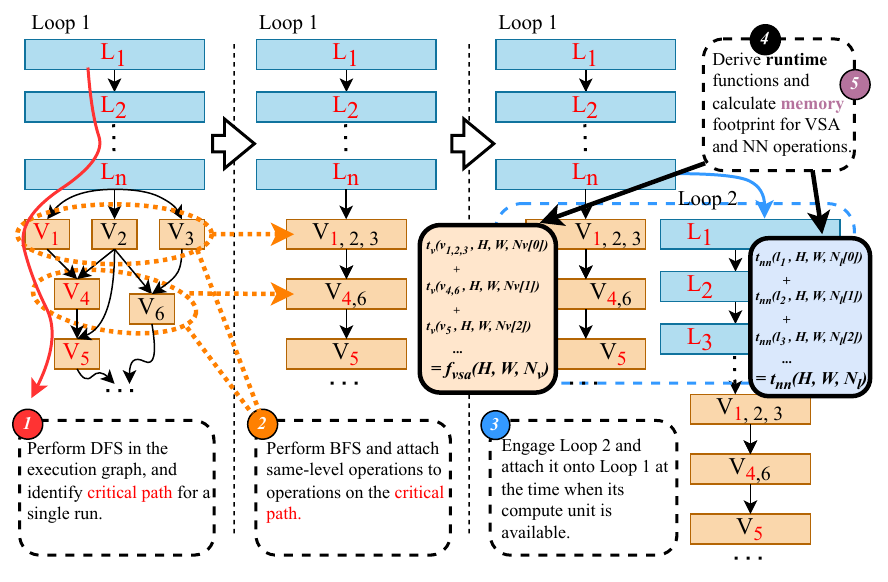}
        \vspace{-20pt}
        \caption{\textbf{Dataflow Architecture Generation (DAG) Flow.}
        }
        \label{fig:dataflow}
        \vspace{-5pt}
\end{figure}

\textbf{Dataflow Graph.} Fig.~\ref{fig:dataflow} depicts how a \textit{Dataflow Graph} is derived. \circled{1} \textbf{Critical path identification:} It begins with a DFS through the execution graph previously generated, identifying the critical path for a single loop of the workload. \circled{2} \textbf{Inner-loop parallelism identification:} Then \DSE walks through the graph again with BFS, to identify operation nodes at the same depth as the nodes on the critical path, and attach them to the corresponding critical-path nodes, indicating their earliest execution and parallelisms. For a single loop, NN layers are typically on the critical path without any attached nodes due to its strict dependencies between layers, while symbolic parts may have more parallelism opportunities thus more grouped nodes. 
\circled{3} \textbf{Inter-loop parallelism identification:} After reshaping the graph for a single loop, \DSE attaches the next loop's graph to the existing one by positioning the first operation of the second loop at the time its required computing unit is freed. For example in the third step shown in Fig.~\ref{fig:dataflow}(b), the first NN layer ($L_1$) in Loop 2 starts as soon as the last NN layer ($L_n$) of Loop 1 finishes and runs along with the symbolic operations in Loop 1. \circled{4} \textbf{Runtime function derivation:} For each node in the newly fused graph, \DSE derives their runtime functions (Sec.~\ref{subsec:dse}) on corresponding sub-arrays with operation parameters (i.e. vector quantity $n$ and dimension $d$, NN layer dimensions in $m, n, k$, etc.) and configuration variables (i.e. sub-array's height, width and partition scheme). \circled{5} \textbf{Memory cost calculation:} \DSE also computes memory footprint based on each node's data size for later memory block configuring.

\textit{Dataflow Graph} describes data dependencies, inner/inter-loop parallelisms as well as their runtime and memory cost model for real-time execution on AdArray. Next we discuss how \DSE uses it to explore the HW design and mapping.


\subsection{Two-Phase Design Space Exploration}
\label{subsec:dse}

\textbf{Design Space.} NSFlow's adaptive architecture (Sec.~\ref{sec:arch}) and the Dataflow Graph (Sec.~\ref{subsec:data_dependency}) creates a large cross-coupled design space, defined by the hardware configuration with height ($H$), width ($W$) and number ($N$) of the sub-arrays, and the mapping scheme specified by the number of sub-arrays running NN layers and VSA operations (i.e. $N_l[i]$ for layer node $i$ and $N_v[j]$ for VSA node $j$, where ), which could vary for each node in the Dataflow Graph during runtime. Note that $N_l$ and $N_v$ are both vector variables with lengths equal to the number of layer and VSA nodes in a single loop. The total size of this design space reaches $10^{300}$ (Tab.~\ref{tab:design_space}), making brutal force search impractical. Next we present how we derive runtime function for the nodes and our unique DSE algorithm.

\begin{table}[t!]
\scriptsize
\centering
\vspace{-10pt}
\caption{\textbf{\NAME Design Space.} \textnormal{Maximum \#PEs = $2^m$. With \textit{exploration phasing} and \textit{space pruning}, search space is reduced by 100 magnitudes.}}
\vspace{-0.05in}
\renewcommand*{\arraystretch}{1.1}
\setlength\tabcolsep{2.1pt}
\resizebox{\linewidth}{!}{%
\begin{tabular}{|l|lll|l|l|}
\hline
         & \multicolumn{3}{l|}{HW config ($H, W, N$)}         & Array partition and mapping      & Total design space, $m = 10$ \\ \hline
Original & \multicolumn{3}{l|}{$m \times (m + 1) / 2$}      & $(N-1)^k$ for each $N$           & $10^{300}$                    \\ \hline
$DAG$    & \multicolumn{3}{l|}{Phase I: $1/4 \leq H/W \leq 16$} & Phase II: Iter $\times$ \#layers & $10^3$                        \\ \hline
\end{tabular}
}
\label{tab:design_space}
\vspace{-10pt}
\end{table}



\begin{table*}[t!]
\centering
\caption{\textbf{Design Configuration and FPGA deployment.}}
\small
\renewcommand*{\arraystretch}{1.0}
\resizebox{\linewidth}{!}{%
\centering
\setlength\tabcolsep{5pt}

\begin{tabular}{|cl|cc|cc|c|ccc|c|cccccc|c|}
\hline
\multicolumn{2}{|c|}{\multirow{2}{*}{Workloads}} & \multicolumn{2}{c|}{Precision}   & \multicolumn{2}{c|}{AdArray Configuration}                                                                                                                                & \multirow{2}{*}{\begin{tabular}[c]{@{}c@{}}SIMD \\ Size\end{tabular}} & \multicolumn{3}{c|}{\begin{tabular}[c]{@{}c@{}}On-chip \\ SRAM Blocks \\ (BRAM)\end{tabular}} & \multirow{2}{*}{\begin{tabular}[c]{@{}c@{}}On-chip \\ Cache \\ (URAM)\end{tabular}} & \multicolumn{6}{c|}{AMD U250 Utilization}                                                                                                          & \multirow{2}{*}{Frequency} \\ \cline{3-6} \cline{8-10} \cline{12-17}
\multicolumn{2}{|c|}{}                           & \multicolumn{1}{c|}{NN}   & Symb & \multicolumn{1}{c|}{\begin{tabular}[c]{@{}c@{}}Size \\ (H, W, N)\end{tabular}} & \begin{tabular}[c]{@{}c@{}}Default Partition \\ ($\bar{N}_l$ : $\bar{N}_v$)\end{tabular} &                                                                       & \multicolumn{1}{c|}{MemA1, MemA2}          & \multicolumn{1}{c|}{Mem B}        & Mem C        &                                                                                     & \multicolumn{1}{c|}{DSP}  & \multicolumn{1}{c|}{LUT}  & \multicolumn{1}{c|}{FF}   & \multicolumn{1}{c|}{BRAM} & \multicolumn{1}{c|}{URAM} & LUTRAM &                            \\ \hline
\multicolumn{2}{|c|}{NVSA}                       & \multicolumn{1}{c|}{INT8} & INT4 & \multicolumn{1}{c|}{32, 16, 16}                                                & 14 : 2                                                                                   & 64                                                                    & \multicolumn{1}{c|}{2.7 MB, 1.1 MB}        & \multicolumn{1}{c|}{2.7 MB}       & 1.6 MB       & 16.2 MB                                                                             & \multicolumn{1}{c|}{89\%} & \multicolumn{1}{c|}{56\%} & \multicolumn{1}{c|}{60\%} & \multicolumn{1}{c|}{34\%} & \multicolumn{1}{c|}{8\%}  & 24\%   & 272 MHz                    \\ \hline
\multicolumn{2}{|c|}{MIMONet}                    & \multicolumn{1}{c|}{INT8} & INT8 & \multicolumn{1}{c|}{32, 32, 8}                                                 & 6 : 2                                                                                    & 64                                                                    & \multicolumn{1}{c|}{3.4 MB, 1.2 MB}        & \multicolumn{1}{c|}{3.4 MB}       & 2.1 MB       & 20.1 MB                                                                             & \multicolumn{1}{c|}{89\%} & \multicolumn{1}{c|}{44\%} & \multicolumn{1}{c|}{52\%} & \multicolumn{1}{c|}{43\%} & \multicolumn{1}{c|}{10\%} & 20\%   & 272 MHz                    \\ \hline
\multicolumn{2}{|c|}{LVRF}                       & \multicolumn{1}{c|}{INT8} & INT4 & \multicolumn{1}{c|}{32, 16, 16}                                                & 14 : 2                                                                                   & 64                                                                    & \multicolumn{1}{c|}{2.7 MB, 0.96 MB}       & \multicolumn{1}{c|}{2.7 MB}       & 1.4 MB       & 15.5 MB                                                                             & \multicolumn{1}{c|}{89\%} & \multicolumn{1}{c|}{56\%} & \multicolumn{1}{c|}{60\%} & \multicolumn{1}{c|}{31\%} & \multicolumn{1}{c|}{7\%}  & 24\%   & 272 MHz                    \\ \hline
\end{tabular}
}

\label{tab:deployment}
\vspace{-15pt}
\end{table*}

\textbf{Analytical models.}
Inspired by the analytical models from previous research \cite{samajdar2020systematic, cogsys}, we derive runtime functions specifically for NSFlow. Since \SA is a scale-out design with row-level partition, the NN runtime for layer node $i$ can be calculated as:

\vspace{-15pt}
\begin{equation}
\scalebox{0.85}{$
    t_{l}(H, W, N_l[i]) = (2H + W + d_{1, i} - 2) \times \lceil \lceil d_{2, i}/N_l[i] \rceil /H \rceil \times \lceil d_{3, i}/W \rceil
$}
\end{equation}
where $H$ and $W$ are sub-array height and width. $d_{1,i}$, $d_{2,i}$ and $d_{3,i}$ are layer dimensions $m, n, k$. Assuming $R_l$ is a set collecting all layer nodes within a loop, NN total runtime is:
\begin{equation}
t_{nn}(H, W, N_l) = \Sigma_{i}^{R_l}t_l(H, W, N_l[i])
\end{equation}
For a VSA node $j$, runtime for spatial and temporal mapping are respectively:

\vspace{-10pt}
\begin{equation}
    t_{v, spatial}(H, W, N_v[j]) = n_j \times \lceil d_j/(W \times H \times N_v[j]) \rceil \times T
\end{equation}
\vspace{-15pt}
\begin{equation}
    t_{v, temp}(H, W, N_v[j]) = \lceil n_j/W \rceil \times \lceil d_j/H \times N_v[j] \rceil \times T
\end{equation}
\vspace{-15pt}

where $T = 3 \times H + d_j -  1$, $n_j$ and $d_j$ are the vector quantity and size respectively. 
\DSE uses the fastest mapping scheme, so with $R_v$ as the set of all VSA nodes, the total VSA runtime in a single loop is:

\vspace{-15pt}
\begin{equation}
\begin{split}
        t_{vsa}(H, W, N_v) = min(\Sigma_{j}^{R_v}t_{v, temp}(H, W, N_v[j]), \\ 
        \Sigma_{j}^{R_v}t_{v, spatial}(H, W, N_v[j]))
\end{split}
\end{equation}


\setlength{\textfloatsep}{2pt} 
\begin{algorithm}[t!]
\label{algo}
\footnotesize
\SetAlgoLined
\KwData{$R_l$, $R_v$, $Range_H$ ($H$ search range), $Range_W$ ($W$ search range), $M$ (max \#PEs), $Iter_{max}$ (Phase II max iterations)}
\KwResult{$H$, $W$, $N$ (total \#sub-arrays), $N_l$, $N_v$}

\BlankLine
\textit{\blue{$/*Phase \ I*/$}}

\For{$H$ in $Range_H$, $W$ in $Range_W$}{
    $N = \lfloor M / (H \times W) \rfloor$ \blue{// get total \#sub-arrays}
    
    \For{$\bar{N_l}$ in $[1, \ N)$}{ 
        \blue{// get optimal HW config for parallel mapping}
        
        Set all elements in $N_l$ to $\bar{N_l}$
        
        Set all elements in $N_v$ to $N - \bar{N_l}$

        $t_{para} = max(t_{nn}(H, W, N_l), t_{vsa}(H, W, N_v))$ 

        \textit{Save the $H, W, \bar{N_l}$ (and $\bar{N_v}$) with minimal $t_{para}$.}

        
    }
    \blue{// get sequential runtime}

    $t_{seq} = \Sigma_{i}^{G}f_{l_i}(H, W, N) + min(\Sigma_{j}^{G}f_{v_j, temp}(H, W, N), \ \Sigma_{j}^{G}f_{v_j, spatial}(H, W, N))$

    \blue{// Set to sequential mode in case it has better performance}
    
    Return and set sequential mode \textbf{if} $t_{seq} < t_{para}$ \textbf{else} Continue
}

\BlankLine
\textit{\blue{$/*Phase \ II*/$}}

\For{$it$ in $Iter_{max}$}{

    \For{layer $i$ in $R_l$}{
        Locate VSA node $j'$ and $j''$ where layer $i$ starts and ends

        \textbf{if} $t_{seq} < t_{para}$ \textbf{do}
            $N_l[i] --$; $N_v[j':j''] ++$;
        
        \textbf{else do}
            $N_l[i] ++$; $N_v[j':j''] --;$
                
        
        $t_{para} = max(t_{nn}(H, W, N_l), \ t_{vsa}(H, W, N_v))$ 

        \textit{Save the $H, W, N_l, N_v$ with minimal $t_{para}$.}
    }
}
    
Return $H$, $W$, $N$, $N_l$, $N_v$.

\caption{\NAME Two-Phase DSE Algorithm}

\label{alg:example2}

\end{algorithm}


\textbf{\SA Design Generation.} We describe how \DSE generates \SA design and mapping scheme in Algorithm~\ref{algo}. To mitigte the search space, the DSE process is decoupled into 2 phases, exploiting AdArray's two-level flexibility (Sec.~\ref{subsec:array}):

In \underline{Phase 1}, \DSE assumes static partitions among nodes to limit search space(i.e. $\forall i, j, N_l[i] = \bar{N_l}$, $N_v[j] = \bar{N_v}$). It finds the optimal $H$, $W$, $N$ constrained by maximum number of PEs $M$ defined based on FPGA resource, and a fixed partition scheme defined by $\bar{N_l}$ and $\bar{N_v}$ to maximizes overall performance. The search space of $H, W, N$ is also pruned based on analytical model results. \underline{Phase 2} explores the mapping scheme further by efficiently fine-tuning $N_l$ and $N_v$ around $\bar{N_l}$ and $\bar{N_v}$ in the dataflow loop with maximum number of iterations pre-defined as $Iter_{max}$, seeking optimal or near-optimal partitions. Searching granularity is set to the span of each NN layer, as VSA kernels are in general smaller and more flexible to be fit into arbitrary array shapes. With $O(N)$ complexity, our two-phase DSE algorithm shrinks the design space by $10^{100}\times$ shown in Tab.~\ref{tab:design_space}.

\begin{table}[t!]
\scriptsize
\centering
\caption{\textbf{\NAME Algorithm Optimization Performance.} \textnormal{\NAME exhibits comparable reasoning capability with the proposed mixed precision.}}
\renewcommand*{\arraystretch}{1.1}
\setlength\tabcolsep{2.1pt}
\resizebox{\linewidth}{!}{%
\begin{tabular}{c|c|c|c|c|c}
\hline
\begin{tabular}[c]{@{}c@{}}Reasoning Accuracy\end{tabular} & FP32 & FP16 & INT8 & MP (IN8 for NN, INT4 for Symb) & INT4 \\ \hline
RAVEN~\cite{raven}                                                      &  98.9\%    &  98.9\%    &   98.7\%   &    98.0\%      &  92.5\%    \\ \hline
I-RAVEN~\cite{iraven}                                                   &  99.0\%    &   98.9\%   &   98.8\%   &     98.1\%     &  91.3\%    \\ \hline
PGM~\cite{barrett2018measuring}                                                         &  68.7\%    &   68.6\%   &   68.4\%  &  67.4\%        &  59.9\%    \\ \hline
Memory                                                    &   32MB   &  16MB    &   8MB   &   5.5MB       &   4MB   \\ \hline
\end{tabular}
}
\label{tab:design_space}
\end{table}

\textbf{Memory and SIMD unit.} After generating the design of \SA, memory blocks sizes are computed based on node memory cost to eliminate inner-node memory stalls, for example $M_{A1}=max(filter \ size \ in \ R_l), \ M_{A2}=max(node \ size \ in \ R_v)$.
$M_{A1}$ and $M_{A2}$ are merged for non-parallel operations.
On-chip cache size is $2\times(M_A + M_B + M_C)$. SIMD size is minimized such that latency of concurrent elem-wise/vector reduction operations can be hidden. 


Unlike other DSE work \cite{kao2020gamma, kwon2020maestro} that only focuses on single task mapping on traditional systolic array, \NAME exploits NSAI inter-task and inner-task parallelism opportunities on \SA at both hardware level and mapping level, boosting the performance of versatile NSAI workloads.

\section{Evaluation Results}
\label{sec:eval}


\subsection{Experimental Setup}
\label{subsec:exp_setup}

\textbf{Algorithm setup.} We evaluate \NAME with three state-of-the-art VSA-based NSAI workloads, i.e., NVSA~\cite{hersche2023neuro}, MIMONet~\cite{menet2024mimonets}, and LVRF~\cite{hersche2024probabilistic} on the commonly-used spatial-temporal reasoning datasets - RAVEN~\cite{raven}, I-RAVEN~\cite{iraven}, PGM~\cite{barrett2018measuring}, CVR~\cite{zerroug2022benchmark}, and SVRT~\cite{fleuret2011comparing}.
Following~\cite{hersche2023neuro,menet2024mimonets,hersche2024probabilistic}, we select the training hyperparameters based on the end-to-end reasoning performance on the validation set. 

\textbf{Hardware setup.} We consider several hardware baselines, including TX2, Xavier NX, Xeon CPU, RTX 2080, and ML accelerators (TPU, Xilinx DPU). 
\NAME framework can be deployed on any type of FPGA board. Tab. \ref{tab:deployment} showcases our deployment for 3 algorithms on AMD U250 using Xilinx Vivado and Synopsys Design Compiler. The clock frequency is set to 272MHz. 

\subsection{\NAME Performance} 
\label{subsec:perf}

\textbf{Mixed-precision performance.}
We benchmark NSAI model on three spatial-temporal reasoning datasets to first evaluate the effectiveness of mixed quantization in NSFlow. As shown in Tab.~ \ref{tab:design_space}, we can observe that \NAME mixed precision achieves comparable accuracy with NVSA algorithm~\cite{hersche2023neuro} while with 5.8$\times$ memory footprint savings. Similar results are observed in MIMONet/LVRF on CVR/SVRT datasets. It is also worth noting that neurosymbolic methods consistently achieve improved cognition and reasoning capability than neural network-based methods and surpass human performance.

\begin{figure}[t!]
\centering
\includegraphics[width=\columnwidth]{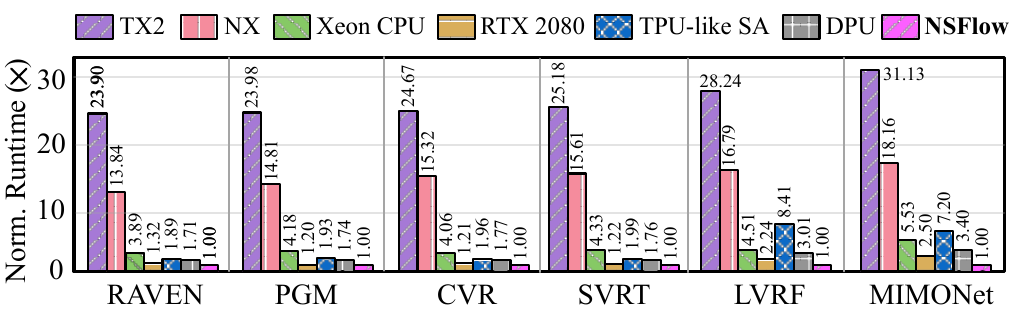}
\vspace{-0.27in}
        \caption{\textbf{End-to-End Runtime Improvement.} \textnormal{\NAME consistently outperforms Xilinx DPU, TPU-like accelerator, Xeon CPU, RTX GPU, and edge SoCs (TX2, NX) in end-to-end runtime evaluated on NSAI reasoning tasks.}}
        \label{fig:runtime}
        \vspace{-5pt}
\end{figure}

\textbf{Performance improvement.} We first benchmark our \NAME accelerator against edge SoC (Jetson TX2, Xavier NX), Intel Xeon CPU, Nvidia RTX 2080, TPU-like systolic array (128$\times$128), and Xilinx DPU. For accelerating NSAI algorithm on six reasoning tasks featuring different difficulties. We can observe in Fig.~\ref{fig:runtime} that \NAME accelerator consistently outperforms other devices, offering 31$\times$/18$\times$ speedup over TX2 and NX, more than 2$\times$ over GPU, up to 8$\times$ speedup over TPU-like systolic array, and more than 3$\times$ speedup over Xilinx DPU on some standard workloads. 

\begin{figure}[t!]
\centering
\includegraphics[width=\columnwidth]{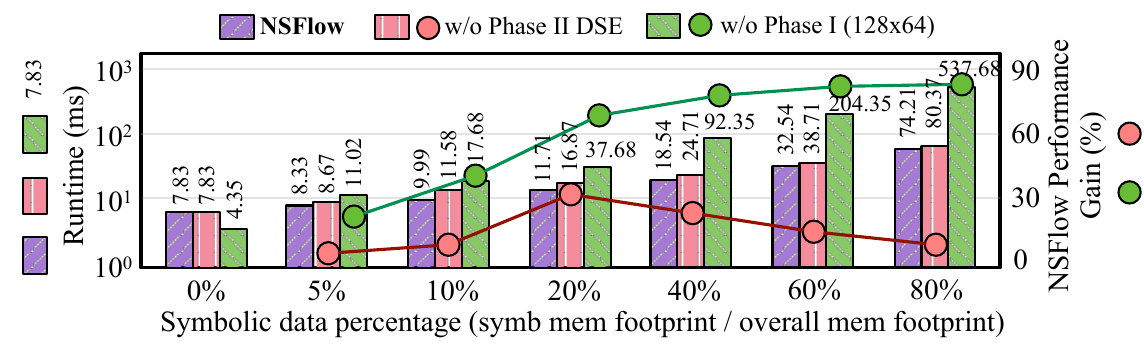}
\vspace{-0.27in}
        \caption{\textbf{Ablation Study.} \textnormal{\NAME exhibits superior scalability comparing to normal TPU design across workloads with various symbolic proportions.}}
        \label{fig:ablation}
\end{figure}

\textbf{Ablation study.} To further showcase the scalability of NSFlow and validate the necessity of proposed DSE algorithm, in Fig.~\ref{fig:ablation} we summarize the runtime of an NSFlow-generated architecture (32$\times$32$\times$8) w/ and w/o the proposed mapping and hardware techniques, evaluated on a NVSA-like workload with varying vector-symbolic data proportions alongside a ResNet18. We observe that despite slight overhead caused by array partition when symbolic part is minimal ($<1\%$), (1) with symbolic ratio going up NSFlow speedup against traditional systolic array grows steadily, reaching up to more than $7\times$ when symbolic data occupies $80\%$ of the memory. (2) The performance gain from our two-phase DSE algorithm (compared to only having array folding, or Phase I) can reach $44\%$ when symbolic workload is balanced with NN (symbolic memory percentage $\approx 20\%$). These findings highlight NSFlow's scalability to adapt to varying workloads and its efficiency in handling symbolic-heavy scenarios.

\vspace{-1pt}
\section{Conclusion}
\label{sec:conclusion}

To enable efficient NSAI for real-time cognitive applications, we propose NSFlow, the first end-to-end design automation framework to accelerate NSAI systems. \NAME leverages the unique NSAI workload characteristics, explores dataflow and architecture design space, and generates scalable designs for FPGA deployment. We believe \NAME paves the way for advancing efficient cognitive reasoning systems and unlocking new possibilities in NSAI.
\bibliographystyle{ieeetr}
\bibliography{refs}

\vspace{12pt}

\end{document}